\documentclass[pre,showpacs,longbibliography,twocolumn]{revtex4-1}
\usepackage{graphicx}
\usepackage{color}
\usepackage{amsmath}
\usepackage{subfigure}
\usepackage{appendix}
\usepackage{soul}
\usepackage{bbm}
\usepackage[colorlinks,linkcolor=blue,citecolor=blue,urlcolor=blue,hyperindex,breaklinks]{hyperref}

\usepackage{float}   
\usepackage{graphicx}

\begin{document}

\title{Quantum thermal diode with additional control by auxiliary atomic states}
\author{Qin Zhang, Zi-chen Zhang, Yi-jia Yang, Zheng Liu, and Chang-shui Yu }
\affiliation{School of Physics, Dalian University of Technology, Dalian 116024, China}
\email{Electronic address: ycs@dlut.edu.cn}
\date{\today}

\begin{abstract}
A quantum thermal diode, similar to an electronic diode, allows for unidirectional heat transmission.
In this paper, we study a quantum thermal diode composed of two two-level atoms coupled to auxiliary two-level atoms.  
We find that the excited auxiliary atoms can weaken heat current and enhance the rectification effect, but the ground-state auxiliary atoms can enhance heat current and weaken the rectification effect. 
The more auxiliary atoms are coupled, the stronger the enhancing or weakening impact is. 
If the auxiliary atom is in a superposition state, we find that only the fraction that projects onto the excited state plays a significant role. 
In particular, if we properly design the coupling of the auxiliary atoms, the rectification effect can be eliminated.
 This provides the potential to control the heat current and the rectification performance by the states of the auxiliary atoms.
\end{abstract}

\maketitle

\section{Introduction}
\label{section1} 
Quantum thermodynamics has attracted increasing interest in recent years and a variety of innovative microscopic thermal devices, including quantum Otto engines \cite{kaur2025performance,PhysRevB.101.054513, Kaur2025,PhysRevA.108.042614,PhysRevResearch.6.023172}, quantum thermometers \cite{PhysRevLett.119.090603, PhysRevB.100.045418,PhysRevLett.128.040502,PhysRevA.109.042417,PhysRevA.111.052216}, quantum refrigerators \cite{Bhardwaj2017, PhysRevApplied.6.054014, PhysRevB.94.235420, PhysRevA.107.032602,PhysRevE.107.044118,PhysRevA.111.012209}, quantum transistors \cite{PhysRevB.105.235412,PhysRevLett.111.063601,PhysRevLett.116.200601,PhysRevE.98.022118,PhysRevE.106.024110,PhysRevB.104.045405,PhysRevB.108.235421,PhysRevB.108.235421}, and quantum switches \cite{Karimi_2017,JAMSHIDIFARSANI20191722,PhysRevLett.121.090503,PRXQuantum.5.010325,10.1063/5.0160675,PhysRevA.107.032412} have been proposed. These devices are not only of great significance in basic research, but also show broad prospects in the application of quantum technology. Quantum thermal diodes, similar to the electric diodes, only allow \cite{RevModPhys.84.1045,PhysRevLett.93.184301,PhysRevLett.93.184301, RevModPhys.84.1045, PhysRevE.95.022128, PhysRevB.98.035414, PhysRevE.99.042102, PhysRevB.99.035129, PhysRevE.99.042121, PhysRevE.104.054137, PhysRevE.103.052130, Mojaveri2021, PhysRevResearch.2.033285, PhysRevE.107.044121, PhysRevE.107.064125, Kasali2022} unidirectional heat transfer significantly affects the energy conversion efficiency, heat rectification, and thermal management performance.

Quantum thermal diodes take advantage of the unique physical properties of quantum systems to achieve efficient rectification functions \cite{Shrestha2020,Luo2021,D1MH00425E}. To improve rectification performance, a variety of strategies have been adopted. For example, adjusting the natural vibration frequency of atoms \cite{RevModPhys.84.1045}, the system can achieve a selective response to heat currents of different frequencies, which shows great significance in nanoscale thermal management devices. The enhancement of coupling strength can significantly improve the performance of quantum thermal diode rectification, especially in the process of interatomic interaction \cite{PhysRevB.103.115413, PhysRevE.89.062109} \cite{PhysRevE.95.022128, PhysRevE.109.014137,10.1063/5.0237842}. The interaction type between atoms
can also affect the rectification ability of the system \cite{PhysRevE.104.054137,DZYALOSHINSKY1958241,PhysRev.120.91}.  Recent studies have shown that common heat reservoirs can enhance the performance of quantum thermal devices and provide additional cross-dissipation channels in some cases \cite{PhysRevA.81.012105, PhysRevA.83.052110, PhysRevA.85.062323,
PhysRevA.97.052309,Hu2018}. This could also have important implications for designing quantum thermal diodes.

\begin{figure}[tbp]
\centering
\includegraphics[width=8.3cm]{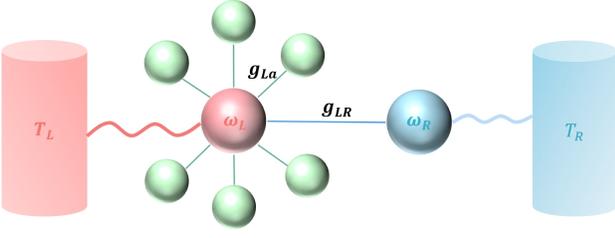} 
\caption{Schematic diagram of the quantum thermal diode. Quantum thermal
diodes consist of two types of atoms: $L$ atoms and $R$ atoms. The $L$ atom
is surrounded by auxiliary atoms, and the $L$ atom and the $R$ atom are
connected to their respective heat reservoirs, temperature $T_L$ and $T_R$.
The coupling strength between the L atom and the R atom is $g_{LR}$, while
the coupling strength between the $L$ atom and the $a$th auxiliary atoms
around it is $g_{La}$.}
\label{M}
\end{figure}

In this paper, we design a quantum thermal diode based on the two-atom system  further coupled to the auxiliary two-level atoms. This system includes the left and the right coupled two-level atoms separately in contact with a heat reservoir.  The left atom interacts with other auxiliary atoms as shown in Fig. {\ref{M}}. The rectification of steady-state heat current is studied based on the global master equation. We find that the states of the auxiliary atoms significantly affect the heat
rectification of the system. When the frequency of the left atom is larger than that of the right atom, the rectification effect reaches its best if all the auxiliary atoms are in the excited states, and reaches its worst if the auxiliary
atoms are in the ground states. When the frequency of the left atoms is lower than that of the right atoms, the result is reversed. The advantage of auxiliary atoms is maintained even though they are weakly dissipative. By regulating the
coupling strength and state distribution of the atoms, the rectification performance of the system can be further optimized, which provides a new possibility for the design of highly efficient energy transmission and rectification devices.  

The article is organized as follows. Section \ref{section2} describes in detail our model and the steady-state dynamics. Section \ref{section3} elaborates on the heat current.  The rectification is discussed in Sec.\ref{section4}. Finally,  conclusions are presented in Sec. \ref{section5}. Some details of the derivations are in the appendix.

\section{Model and steady state}
\label{section2} 

Our system consists of two coupled atoms, each connected to its respective
thermal reservoir with temperature $T_L$ and $T_R$ respectively.  One of the atoms (referred to as $L$ atom) interacts with $N$ additional auxilliary
atoms (labeled as atom $1, 2,\cdots, N$). 
The Hamiltonian of the dissipative system reads (we take $\hbar=k_B=1$) 
\begin{align}
&H_S=H_{S0}+H_{SI},\\
&H_{S0}=\frac{1}{2} \omega_{L} \sigma^{z}_{L}+\frac{1}{2} \omega_{R}
\sigma^{z}_{R}+\sum_{a=1}^{N}\frac{1}{2}\omega_{a} \sigma^{z}_{a},  \notag \\
&H_{SI}=g_{LR}\sigma_L^z\sigma_R^z+\sum_{a=1}^{N}g_{La}\sigma_L^z\sigma_a^z,\notag
\end{align}
where $\omega_\mu$ represent the
transition frequencies of atoms, $\sigma^{z}_{\mu}$ ($\mu=L,R,a$) is the  Pauli matrix, $%
g_{L\mu}$ denotes the coupling strength between the atom $L$ and the atom $%
\mu$th.  The left and right atoms are respectively connected to their harmonic
oscillator heat reservoirs with temperatures $T_L$ and $T_R$, while the auxiliary atoms are not connected to the heat reservoirs. Werlang et al. \cite{PhysRevE.89.062109} studied the thermal transport mechanism of a two-spin system and its respective boson heat reservoir under strong coupling. In this paper, we also adopt the boson heat reservoir.
The free Hamiltonian of the two thermal reservoirs is 
\begin{align}
H_E=\sum_k\omega_{L k}b_{L k}^\dagger b_{L k}+\sum_k\omega_{R k}b_{R
k}^\dagger b_{R k},
\end{align}
where $\omega_{\mu k}$ is the frequency of reservoir $\mu$ at the $k$th
Bosonic mode and $b_{\mu k}^\dagger$ and $b_{\mu k}$ represent the bosonic creation and annihilation operators for the modes of the reservoir. Ref.\cite{Ghosh_2022} used the fermion heat reservoir.
The interaction between the atoms and their respective
thermal reservoirs are given by \cite{CALDEIRA1983374} 
{\begin{align}
H_{SE}=\sum_kf_{Lk}\sigma_L^{-} b_{L k} +\sum_kf_{R
k}\sigma_R^{-} b_{R k}+\mathrm{h.c.},
\end{align}
where $\sigma_{L(R)}^-$ represents the lowering operator, $f_{\mu k}$ represents the coupling strength between the $\mu$ atom
and the $k$th bosonic mode.

To study the dynamics of the system, we will derive the master equation based on the
Born-Markov-secular approximation \cite{GORINI1978149, 10.1093/acprof:oso/9780199213900.001.0001, doi:10.1142/6738, PhysRevA.84.043832,PhysRevA.108.042212}. 
Due to the $z$-$z$ coupling in the system \cite{2019Analysis}, the
Hamiltonian takes the form of a diagonal matrix. 
Thus, the eigenstates of the system can be written as the direct product of each atom's eigenstate. The
system consists of $N+2$ atoms,  so there are $2^{N+2}$ eigenstates $%
\vert i\rangle$, $i$=[1,$2^{N+2}$], which can be explicitly given as $\left\vert 1\right\rangle$=$%
\left\vert e  e  e \cdots e  \right\rangle$, $\left
\vert 2\right\rangle=\left\vert eee\cdots g\right \rangle$, $\cdots
$, $\left\vert 2^{N+2}\right\rangle$=$\left\vert ggg\cdots%
g\right\rangle$, where $\left\vert e  e  e \cdots e %
\right\rangle=\left\vert e \right\rangle_L\otimes\left\vert e \right\rangle_R\otimes\left\vert%
 e \right\rangle_1\otimes\cdots\otimes\left\vert e \right\rangle_N$.
Thus, one can obtain  the eigen-operator $V_{\mu
l}\equiv V_{\mu}(\omega_{\mu l})=\sum_{\omega_{\mu l}=E_j-E_i}\vert i\rangle\langle i
\vert\sigma_{\mu}^x\vert j\rangle\langle j \vert$, where $E_{i}$ is the eigenvalue
of the system, and $l$ marks the different frequencies.  After a simple calculation, the left and right atoms have $%
2^{N+1}$ and $2$ eigenoperators respectively, which, as well as  the corresponding
eigenfrequencies,  are explicitly given as $\sigma_L^-\otimes\rho_R^{\pm}\otimes%
\rho_1^{\pm}\otimes \cdots \otimes\rho_N^{\pm}$, $\omega_L\pm2g_{LR}\pm2
g_{L1}\pm \cdot\cdot\cdot \pm2g_{LN}$, $\rho_L^\pm\otimes\sigma_R^-\otimes\mathbbm{I}_1\otimes \cdots \otimes\mathbbm{I}_N$, $\omega_R\pm2g_{LR}$. $%
\sigma^-_{\mu}=\vert g\rangle_{\mu}\langle e \vert$ represents
the lowering operator of the $\mu$ atom, $\rho^+_{\mu}=\left\vert e \right\rangle_{%
\mu}\left\langle  e \right\vert$ and $\rho^-_{\mu}=\left\vert g\right\rangle_{\mu}%
\left\langle g\right\vert$ are the excited and the ground states of  the $\mu$th atom, and $%
\mathbbm{I}_{a}$ is the two-dimensional identity matrix of the $a$th atom.
The transition frequency of the atoms on the right is only
affected by $g_{LR}$, while the transition frequency of the atoms on the
left is affected by both $g_{LR}$ and $g_{La}$.

Based on the eigen-operators, in accordance with standard procedures\cite{10.1093/acprof:oso/9780199213900.001.0001}, one can get the master equation in
 Schr$\ddot{\mathrm{o}}$dinger picture as
\begin{align}
\dot{\rho}(t)=-i[H_S,\rho(t)]+\mathcal{L}_L[\rho(t)]+\mathcal{L}_R[\rho(t)],
\label{dissipator}
\end{align}
where the Lindblad dissipators $\mathcal{L}_L[\rho(t)]$ and $\mathcal{L}%
_R[\rho(t)]$ corresponding to the $L$th and $R$th atom are defined as 
\begin{align}
\mathcal{L}_L[\rho(t)]&=\sum_{l=1}^{2^{N+1}}J_L(-\omega_{Ll})[2V_{Ll}\rho (t)%
{V_{Ll}}^\dagger-\{{V_{Ll}}^\dagger V_{Ll},\rho(t)\}]  \notag \\
&+J_L(+\omega_{Ll})[2{V_{Ll}}^\dagger\rho (t){V_{Ll}}-\{{V_{Ll}V_{Ll}}%
^\dagger,\rho(t)\}] ,  \notag  \label{dissipatormu} \\
\mathcal{L}_R[\rho(t)]&=\sum_{l=1}^{2}J_R(-\omega_{Rl})[2V_{Rl}\rho (t){%
V_{Rl}}^\dagger-\{{V_{Rl}}^\dagger V_{Rl},\rho(t)\}]  \notag \\
&+J_R(+\omega_{Rl})[2{V_{Rl}}^\dagger\rho (t){V_{Rl}}-\{{V_{Rl}V_{Rl}}%
^\dagger,\rho(t)\}].
\end{align}
In Eq. (\ref{dissipatormu}) $J_{\mu} (\pm\omega_{\mu
l})=\pm\gamma_{\mu}(\omega_{\mu l})n_{\mu}(\pm\omega_{\mu l})$ represents
the spectrum density {\cite{PhysRevE.85.061126} and $n_{\mu}(\omega_{\mu l})=[\mathrm{exp}(\omega_{\mu
l}/T_{\mu})-1]^{-1}$ is the average photon number of mode $\omega_{\mu l}$
at temperature $T_{\mu}$. The dissipation rate $\gamma_{\mu}(\omega_{\mu l})$
between the $\mu$th atom and the corresponding reservoir with the frequency $%
\omega_{\mu l}$ is a square relationship with the spin-boson coupling
strength $f_{\mu}(\omega_{\mu l})$, i.e. $\gamma_{\mu}(\omega_{\mu l})=\pi
f_{\mu}^2(\omega_{\mu l})$. To simplify our analysis, we assume a flat spectrum. Under this assumption, the frequency
distribution of the reservoir is uniform, i.e. $\gamma_{\mu}(\omega_{\mu
l})=\gamma_{\mu}=\gamma$.    

The dynamics of the density matrix Eq. (\ref%
{dissipator}) can be divided into  the diagonal part and
the off-diagonal part. The diagonal part represents the dynamics of the populations, while
the off-diagonal part reflects the coherence \cite{PhysRevE.76.031115}. 
In most cases, the dynamics of the off-diagonal entries give $%
\dot{\rho}_{i,j}=-\eta\rho_{i,j}$,  where $\eta$ is a real number related to the
spectral density.  For $i \in [1,2^N-1], j \in [i+1,2^N]$, the dynamics of the off-diagonal entries are explicitly given in Appendix \ref{Appendix A}.  Without loss of generality,  we take $%
\dot{\rho}_{i,j}$ and $\dot{\rho}_{i+2^N,j+2^N}$ as an example.  The dynamical equation can be 
expressed as
\begin{align}
\begin{pmatrix}
\dot{\rho}_{i,j} \\ 
\dot{\rho}_{i+2^{N},j+2^{N}}%
\end{pmatrix}
=\varLambda 
\begin{pmatrix}
\rho_{i,j} \\ 
\rho_{i+2^{N},j+2^{N}}%
\end{pmatrix}%
,
\end{align}
where $\varLambda$ is the coefficient matrix and depends only on the
spectral density. In the steady state, $\dot{\rho}_{i,j}=0$. We can see that
det$(\varLambda) \neq0$, so $\rho_{i,j}=0$. So, in the steady state, the
off-diagonal elements disappear, and so does the coherence.

Now let's turn to the diagonal part. The diagonal elements $%
\rho_{i,i}$ satisfy the equation \cite{PhysRevE.89.062109,PhysRevE.90.052142}
\begin{align}
\dot{\rho}_{i,i}=-\sum_{j=1,j\neq i}\Gamma_{i,j}^L-\sum_{j=1,j\neq
i}\Gamma_{i,j}^R ,\quad i=1,\cdot\cdot\cdot,2^{N+2},  \label{dynamic_total}
\end{align}
where $\Gamma^{\mu}_{i,j}=2[J_{\mu}(-\omega_{\mu l}){\rho}_{i,i}-J_{\mu
}(+\omega_{\mu l})\rho_{j,j}]=-\Gamma^{\mu}_{j,i}$ represents the net
transition rate from state $\vert i\rangle$ to the state $\vert j\rangle$
resulting from interactions with the $\mu$ reservoir. The steady-state solution of
 Eq. (\ref{dynamic_total}) can be obtained by solving
for 
\begin{equation}
\vert\dot{\rho}\rangle=\mathcal{M}\vert\rho\rangle=0,  \label{Condsteady}
\end{equation}
where $\vert\dot{\rho}\rangle$ is a column vector composed of the diagonal elements, and $\mathcal{M}=\mathcal{M}_L+\mathcal{M}_R$ represents the coefficient
matrix \cite{PhysRevE.90.042142}. 

From the  eigenoperator $V_{Ll}$, $V_{Rl}$ and Eq. (\ref{dynamic_total}%
),  one can find that  the four energy levels  $%
\left\vert e_L  e_R\right\rangle\otimes\left\vert\phi_m\right\rangle$, $\left\vert e_L
g_R\right\rangle\otimes\left\vert\phi_m\right\rangle$, $\left\vert g_L
 e_R\right\rangle\otimes\left\vert\phi_m\right\rangle$, $\left\vert g_L
g_R\right\rangle\otimes\left\vert\phi_m\right\rangle$ form an independent subspace in which the transitions are irrelevant of the energy levels out of the subspace,  where  $\left\vert\phi_m\right\rangle$=$\otimes_{a=1}^N
\left\vert\phi_m^a\right\rangle$ represents  $m=2^N$ possible states of the $N$
 auxiliary atoms with $\left\vert\phi_m^a\right\rangle$=$\vert e \rangle_a$(or $%
\vert g\rangle_a$) denote the excited (or the ground) states of the $a$th atom. Thus, the total system is divided into $2^N$ independent subspaces.  In each subspace, the transition frequencies $\omega_{ij}$ between $i$th and $j$th energy levels are related to the eigenfrequency as $\omega_{m,m+2^{N+1}}=\omega_{L,m}$, $%
\omega_{m+2^N,m+2^{N+1}+2^N}=\omega_{L,m+2^N}$, $\omega_{m,m+2^N}=%
\omega_{R,1}$, $\omega_{m+2^{N+1},m+2^{N+1}+2^N}=\omega_{R,2}$.  Thus, the
coefficient matrix and the diagonal-element vector can be written as $\mathcal{M}%
=\left (\bigoplus_{m=1}^{2^N} \mathcal{M}_L^m\right)\otimes\mathrm{I}_R+%
\mathrm{I}_L\otimes\mathcal{M}_R$ and $\vert\rho\rangle$=$%
\sum_{m=1}^{2^N}p_m\vert\phi_m\rangle\otimes\vert\rho_m\rangle$, $%
\sum_{m=1}^{2^N}p_m=1$, where $\vert\rho_m\rangle\equiv[\rho_{\mathit{11}%
},\rho_{\mathit{22}},\rho_{\mathit{33}},\rho_{\mathit{44}}]^T$,
respectively. Here, $p_m$ represents the probability of each independent
subspace, and $\rho_{\mathit{11}}\equiv\rho_{m,m}$, $\rho_{\mathit{22}%
}\equiv\rho_{m+2^N,m+2^N}$, $\rho_{\mathit{33}}\equiv%
\rho_{m+2^{N+1},m+2^{N+1}}$, $\rho_{\mathit{44}}\equiv%
\rho_{m+2^{N+1}+2^N,m+2^{N+1}+2^N}$. For convenience,  let $%
J_{\mu\pm}^{m}\equiv J_{\mu}(\pm\omega_{\mu m})$ , then  
\begin{align}
\mathcal{M}_L^m &= 2 
\begin{pmatrix}
-J_{L-}^m & 0 & J_{L+}^m & 0 \\ 
0 & -J_{L-}^{m+2^N} & 0 & J_{L+}^{m+2^N} \\ 
J_{L-}^m & 0 & -J_{L+}^m & 0 \\ 
0 & J_{L-}^{m+2^N} & 0 & -J_{L+}^{m+2^N}%
\end{pmatrix}%
,  \notag \\
\mathcal{M}_R &= 2 
\begin{pmatrix}
-J_{R-}^1 & J_{R+}^1 & 0 & 0 \\ 
J_{R-}^1 & -J_{R+}^1 & 0 & 0 \\ 
0 & 0 & -J_{R-}^2 & J_{R+}^2 \\ 
0 & 0 & J_{R-}^2 & -J_{R+}^2%
\end{pmatrix}%
.  \label{mathcal_R}
\end{align}
Solving the steady-state equation $\mathcal{M}\vert\rho^s\rangle$=0,
we have
$\rho_{\mathit{ii}}^s=\frac{\tilde{\rho}_{\mathit{ii}}^s}{N}$, $\mathit{i}=[%
\mathit{1}, \mathit{4}]$, where  
\begin{align}
\tilde\rho_{\mathit{11}%
}^s=J_{L+}^{m+2^N}J_{R+}^1(J_{L+}^m+J_{R-}^2)+J_{L+}^mJ_{R+}^2(J_{L-}^{m+2^N}+J_{R+}^1),
\notag \\
\tilde\rho_{\mathit{22}%
}^s=J_{L+}^mJ_{R-}^1(J_{L+}^{m+2^N}+J_{R+}^2)+J_{L+}^{m+2^N}J_{R-}^2(J_{L-}^m+J_{R-}^1),
\notag \\
\tilde\rho_{\mathit{33}%
}^s=J_{L-}^mJ_{R+}^1(J_{L+}^{m+2^N}+J_{R+}^2)+J_{L-}^{m+2^N}J_{R+}^2(J_{L-}^m+J_{R-}^1),
\notag \\
\tilde\rho_{\mathit{44}%
}^s=J_{L-}^{m+2^N}J_{R-}^1(J_{L+}^m+J_{R-}^2)+J_{L-}^mJ_{R-}^2(J_{L-}^{m+2^N}+J_{R+}^1),\label{state}
\end{align}
and the normalized coefficient $N=\tilde\rho_{\mathit{11}}^s+\tilde\rho_{\mathit{%
22}}^s+\tilde\rho_{\mathit{33}}^s+\tilde\rho_{\mathit{44}}^s$.

\section{Heat current}
\label{section3}

\begin{figure*}[t]
	\centering
	  \includegraphics[width=\textwidth]{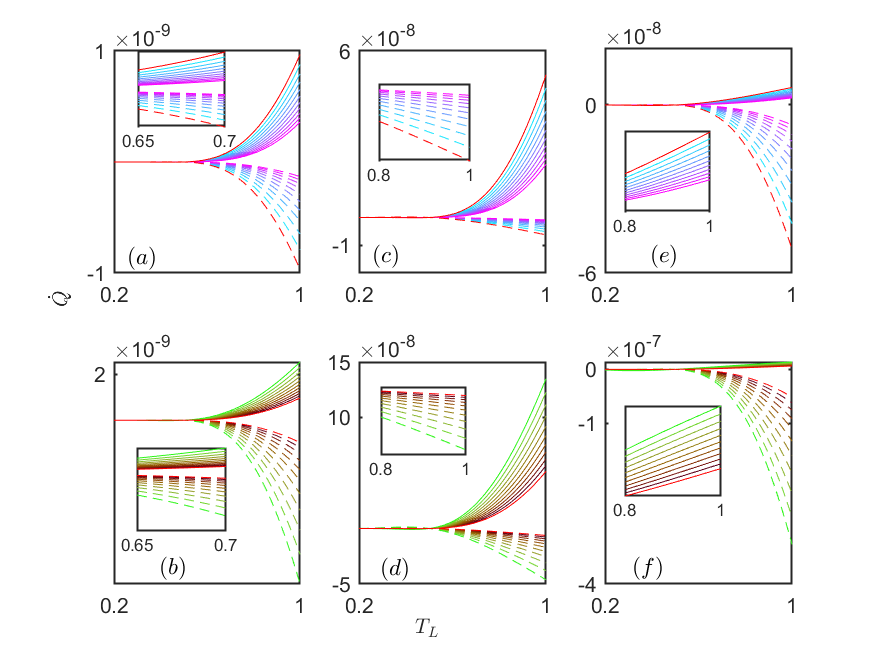}
	\caption{The heat current $\dot{Q}_L$ varies with the temperature $T_L$.  Here, $T_R=0.5$ is fixed. All solid lines represent the positive heat current, and dashed lines represent the reverse heat current by exchanging the temperatures $T_L$ and $T_R$. The red line indicates that the $L$ atom is not connected to any auxiliary atoms.  In (a), (c), and (e),  the auxiliary atoms are at excited states, and the color changes from cyan to magenta, indicating increasing auxiliary atom number  $N$  from $1$ to $10$. In (b), (d), and (f), the auxiliary atoms are at ground states, and the color from brown to green indicates increasing the auxiliary atom number $N$ from $1$ to $10$. In  (a) and (b), the inset shows the positive and reverse heat current over $T_L$ from $0.65$ to $0.7$. In  (c) and (d), the inset shows the  reverse heat current over $T_L$ from $0.8$ to $1$. In  (e) and (f), the inset shows the positive heat current over $T_L$ from $0.8$ to $1$. In  (a) and (b), $\omega_L=4$ and $\omega_R=4$, in (c) and (d), $\omega_L=4$ and $\omega_R=2$, in (e) and (f), $\omega_L=2$ and $\omega_R=4$.  Other parameters are $\omega_a=2$, $g_{LR}=0.1$, $g_{La}=0.05$, $\gamma=0.001$.}
\label{QT}
\end{figure*}

Obtaining the steady-state density matrix $\rho$, we can study the heat current defined by \cite{10.1093/acprof:oso/9780199213900.001.0001}

\begin{align}
\nonumber
\dot{Q}_L=&\mathrm{Tr} \{H_S\mathcal{L}_L[\rho(t)]\},\\
\dot{Q}_R=&\mathrm{Tr} \{H_S\mathcal{L}_R[\rho(t)]\}. \label{Q}
\end{align}
  
 $\dot{Q}_{L(R)}>0$ means that heat current flows from the reservoir to the
system and  $\dot{Q}_{L(R)}<0$ means that the heat current is from the system to the
reservoir. When the system reaches the steady state,  the first law of thermodynamics \cite{PRXQuantum.2.030202}  
$\dot{Q}_L+\dot{Q}_R=0$ is satisfied.  Considering our current system, one can find that 
heat current can be expressed as the sum of the branch heat current of each independent subspace, as
\begin{align}
\dot{Q}_{L}&=\sum_{m=1}^{2^{N}}\dot{Q}_{L,m}=-\sum_{m=1}^{2^N}p_m
4g_{LR}\Gamma_{m,m+2^{N+1}}^L,  \notag \\
\dot{Q}_{R}&=\sum_{m=1}^{2^{N}}\dot{Q}_{R,m}=\sum_{m=1}^{2^N}p_m
4g_{LR}\Gamma_{m,m+2^{N}}^R.  \label{Q1} 
\end{align} 
Substituting Eq. (\ref{state}) into Eq. (\ref{Q1}),  one can find that the net transition rates in each independent subspace are equal, i.e.,
\begin{align}
\nonumber
&\Gamma_{m,m+2^{N+1}}^L=-\Gamma_{m+2^N,m+2^{N+1}+2^N}^L\\
=&\Gamma_{m,m+2^{N}}^R=-\Gamma_{m+2^{N+1},m+2^{N+1}+2^{N}}^R\equiv\Gamma_m^N
\end{align}
with
\begin{align}
\nonumber
\Gamma_m^N&=\frac{1}{N}[J_L(-\omega_{L,m})J_L(\omega_{L,m+2^{N+1}})J_R(\omega_{R,1})J_R(-\omega_{R,2})\\
&-J_L(\omega_{L,m})J_L(-\omega_{L,m+2^{N+1}})J_R(\omega_{R,2})J_R(-\omega_{R,1})].
\end{align}  
See the derivation process in Appendix {\ref{Appendix B}}. During each cycle, the left reservoir exchanges energy at a rate of $|\Gamma_m^N|$, absorbing the amount $\omega_{m,m+2^{N+1}}$ of energy and releasing $\omega_{m+2^{N+1}+2^N,m+2^N}$ of energy\cite{PhysRevE.89.062109}. 

In FIG. \ref{QT}, we plot that the heat current $\dot{Q}_L$ varing with temperature $T_L$.  Note that the dashed lines denote the reverse heat current obtained by exchanging the temperatures $T_L$ and $T_R$. In (a), (c), and (e),  one can find that as the number of auxiliary atoms at excited states increases,  the heat current decreases. In (b), (d), and (f), as the number of auxiliary atoms at ground states increases,  the heat current increases.
In addition, Eq. (\ref{Q1}) implies that increasing the number of auxiliary atoms, the heat current could remain unchanged if the increment of the heat current is
$\sum_{m=1}^{2^N-1}p_m\triangle\dot{Q}_{L,m}=\dot{Q}_L^\prime-\dot{Q}_{L,2^N}$%
, where 
\begin{align}
\dot{Q}_L^\prime=-4g_{LR}\Gamma_{0}^L \label{Q5}
\end{align}
 is the steady-state heat current
without auxiliary atoms and $\triangle\dot{Q}_{L,m}$ denotes the difference between the heat current of the $m$th independent subspace and
that of the $2^N$th independent subspace.

One can find that the heat current only depends on the fraction of the excited states, regardless of the coherence of the auxiliary atoms. To demonstrate this, we consider adding only one auxiliary atom without loss of generality.  Let the $L$ and $R$ atoms be initially in
the excited states and 
 introduce an auxiliary atom in two cases. One is that the auxiliary atom is in a superposition state $%
\left\vert\psi\right\rangle_0=\vert e\rangle_L\otimes\vert e\rangle_R\otimes(\alpha\vert%
e\rangle_1+\beta\vert g\rangle_1)$ with $\alpha$ being a complex
number, and the other case is that the atom is in the mixed state $\rho_0=\vert e \rangle_L\langle e \vert \otimes \vert  e\rangle_R \langle e\vert \otimes (p_1\left\vert e\right\rangle_1\left\langle e\right\vert+(1-p_1)\left\vert g\right\rangle_1\left\langle g\right\vert)$.  This state indicates that the system has coherence at the initial time. 
By re-solving Eq.(\ref{dissipator}), the density matrix of the system is obtained, and substituting it into Eq.(\ref{Q}), the evolution law of heat current $\dot{Q}_L$ with respect to time $t$ can be further derived.
We plot the evolution of heat current over time in  FIG. {\ref{t}}.  It shows that different $\alpha$ or $p_1$ correspond to different steady-state heat currents. As time goes by, when the system reaches the steady state, the heat currents are the same if $\left\vert\alpha\right\vert^2=p_1$.  It indicates that as the system evolves to the steady state, the off-diagonal elements of the density matrix disappear, coherence is lost, and only the diagonal elements remain.

\begin{figure}[t]
	\centering
		\includegraphics[width=8.3cm]{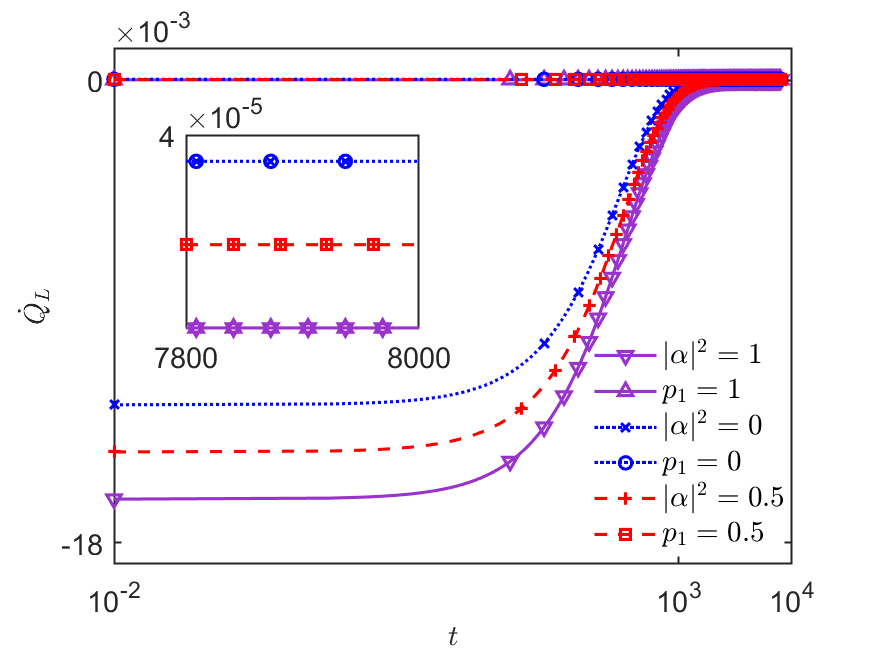}
	\caption{The heat current $\dot{Q}_L$ varies with time $t$. The purple solid line with inverted triangles, the blue dotted line with crosses, and the red dashed line with plus signs, correspond to the state $\left\vert\psi\right\rangle_0$.  In contrast, the purple solid line with regular triangles, the blue dotted line with circles, and the red dashed line with squares, correspond to $\rho_0$. The inset shows the heat current over time $t$ from $7,800$ to $8,000$. The values of $\vert\alpha\vert^2$ and $p_1$ are given in the figure. Other parameters are $\omega_L=5$, $\omega_R=3$, $\omega_1=2$, $\gamma=0.001$, $g_{LR}=1$, $g_{L1}=0.5$, $T_L=2$, $T_R=1$.}
\label{t}
\end{figure}

\section{rectification}
\label{section4} 
Next, we will study the effect of auxiliary atoms on rectification performance. 
The rectification factor $\mathcal{R}$ is defined as  \cite{PhysRevLett.102.095503}
\begin{equation}
\mathcal{R}=\frac{\vert\dot{Q}_L^f+\dot{Q}_L^r\vert}{\max[\vert\dot{Q}%
_L^f\vert,\vert\dot{Q}_L^r\vert]},  \label{R}
\end{equation}
where $\dot{Q}_L^f$ represents the positive heat current for a given $T_L$
and $T_R$, $\dot{Q}_L^r$ represents reverse heat current exchange $T_L$ and $%
T_R$.  $\mathcal{R}=1$ means that the system
is a perfect diode;  $0<\mathcal{R}<1$ indicates that the system exhibits a good
diode;  $\mathcal{R}=0$ means no rectification effect.
Substituting Eq. (\ref{Q1}) into Eq. (\ref{R}),  we have
\begin{align}
\mathcal{R}=1-\frac{A(T_L,T_R)}{A(T_R,T_L)}  \label{R1}
\end{align}
with 
\begin{align}
\nonumber
A(T_L,T_R)&={\cosh}(\frac{2g_{LR}}{T_R})+{\cosh}(\frac{\omega_{R}}{T_R})+{\cosh}(\frac{2g_{LR}}{T_L})\\
\nonumber
&+{\cosh}(\frac{\omega^\prime_{L}}{T_L})+{\sinh(\frac{2g_{LR}}{T_R}){\sinh}\frac{\omega^\prime_L}{T_L})}\\
&+{\sinh}(\frac{2g_{LR}}{T_L}){\sinh}(\frac{\omega_{R}}{T_R}).\label{aa}
\end{align} 
Here, $A(T_R,T_L)$ represents the exchange of $T_L$ and $T_R$, $\omega_L^\prime=\omega_L\pm\sum_{a=1}^N
2g_{La}$ with $\text{``+''}$ corresponding to all the auxiliary atoms in excited states and $\text{``-''}$ to the ground states. The condition for Eq. (\ref{R1}) to hold is that the numerator is smaller than the denominator; otherwise, the two terms need to be interchanged. The derivation can be found in Appendix {\ref{Appendix B}}. From Eq. (\ref{aa}), one can find that  $\omega_L^\prime=\omega_R$ can lead to $A(T_L,T_R)=A(T_R,T_L)$, i.e. $\mathcal{R}=0$. In this sense, $\omega_L^\prime$ can be understood as the effective frequency of a single virtual left atom that collects all the left atoms as a whole.
\begin{figure}
\centering
\includegraphics[width=8.3cm]{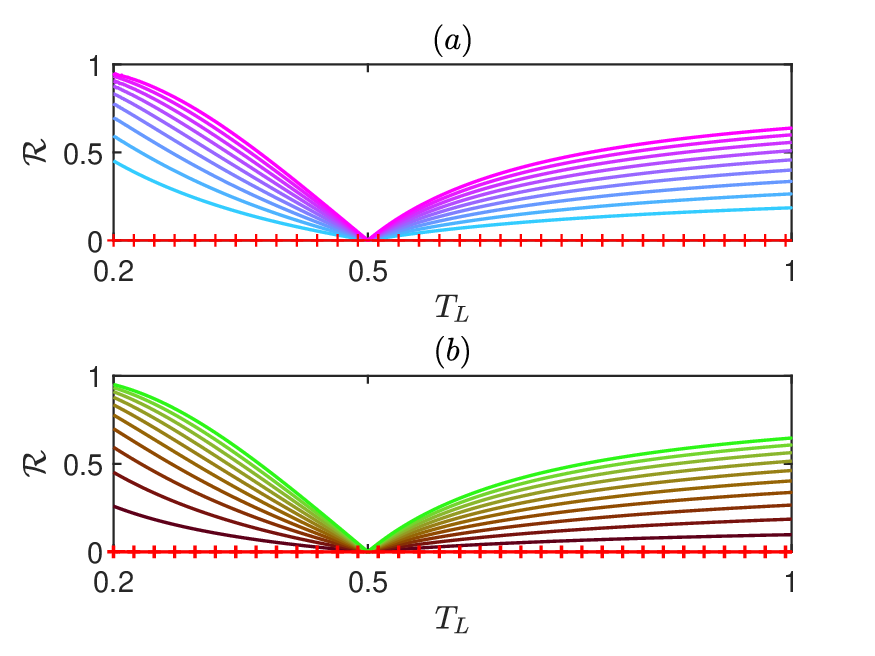} 
\caption{The rectification $\mathcal{R}$ as a function of the temperature $T_L$. The red line with plus signs indicates the absence of auxiliary atoms,
 while the solid line indicates the presence of auxiliary atoms. From bottom to top, $N$ goes from $1$ to $10$. 
 In (a), the auxiliary atoms are all at excited states, and in (b), they are all at ground states. In this figure, $\protect\omega_L=\protect%
\omega_R=4$, $\protect\omega_a=2$, $g_{LR}=0.2$, $g_{La}=0.05$, $\protect%
\gamma=0.001$, $T_R=0.5$.}
\label{D}
\end{figure}

In Fig. {\ref{D}}, we plot the rectification factor as a function of temperature with different auxiliary atoms for $\omega_L=\omega_R$.  The red line with plus signs represents the case without auxiliary atoms, indicating that the system exhibits no rectification effect. When auxiliary atoms are added, it can be observed from Fig. {\ref{D}} that the rectification effect of the system is significantly enhanced with the increase of the
number of auxiliary atoms, regardless of whether the auxiliary atoms
are excited or not. This is consistent with our understanding. 
Namely, although the frequencies of the left and right atoms are the same, as the number of auxiliary atoms increases, the system gradually becomes
asymmetrical, which makes the rectification effect more significant.

The rectification effect emerges in the $\omega_L\neq\omega_R$ system even without auxiliary atoms, while the introduction of auxiliary atoms influences the rectification effect through their states. In Fig. {\ref{RT}%
}, we plot the rectification factor $\mathcal{R}$ with $T_L$. One can find that the number of excited atoms can enhance the rectification effect for $\omega_L>\omega_R$, but the ground-state atom number will weaken the rectification. For $\omega_L<\omega_R$, the role of auxiliary atoms is opposite to the case of $\omega_L>\omega_R$. 
\begin{figure}
\centering
\includegraphics[width=8.3cm]{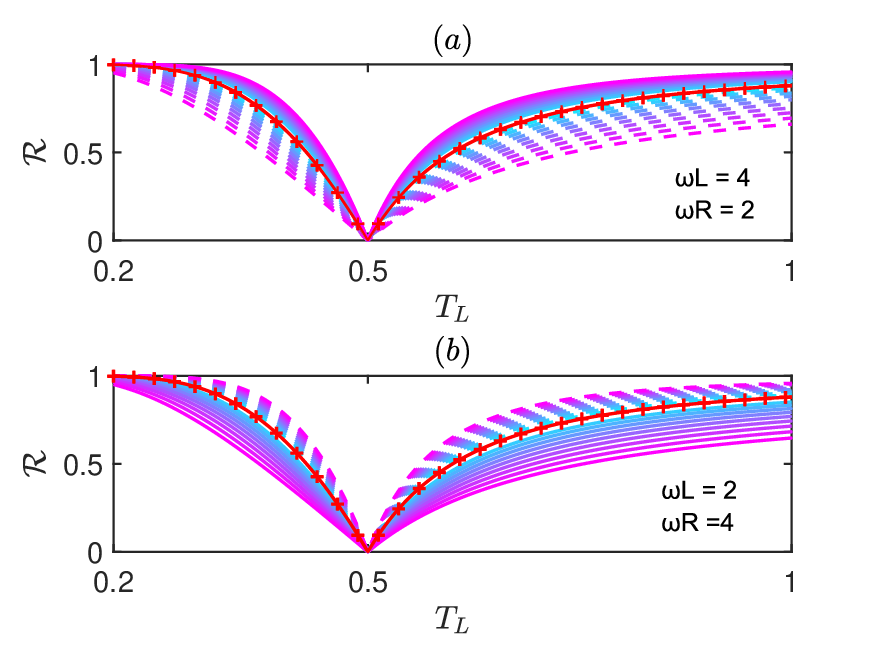} 
\caption{The rectification $\mathcal{R}$ varies with the temperature $T_L$. The red solid line with plus signs indicates the 
absence of the auxiliary atoms, while the solid and dashed
lines indicate the presence of the auxiliary atoms. The solid line
represents all the excited auxiliary atoms, and the dashed line
represents the ground-state auxiliary atoms. The color from cyan to magenta indicates an increase
in the number of atoms $N$, from 1 to 10. Here we take $\protect\omega_a=2$, $g_{LR}=0.1$, $%
g_{La}=0.05$, $\protect\gamma=0.001$, $T_R=0.5$.}
\label{RT}
\end{figure}

\begin{figure}[t]
\centering
\includegraphics[width=8.3cm]{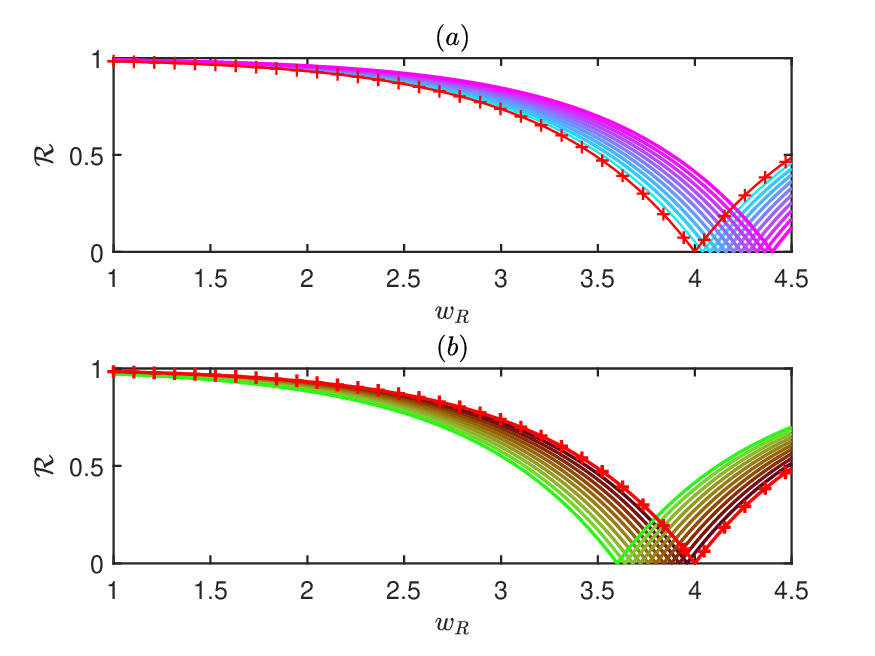} 
\caption{The rectification factor $\mathcal{R}$ varies with the frequency of the
atom on the right $\protect\omega_R$. The lines and colors in the figure
are defined the same as in Fig. {\ref{D}}. Here, $\protect\omega_L=4$, $\protect\omega_a=2$, $\protect\gamma=0.001
$, $g_{LR}=0.2$, $g_{La}=0.02$, $T_L=0.3$, $T_R=0.5$.}
\label{W}
\end{figure}
From Fig. {\ref{RT}}, we can observe that the
rectification factor is closely related to the frequencies of $L$ and $R$
atoms. Next, we will delve into the impact of this frequency on
rectification performance.  In the Fig. {\ref{W}}(a),
when all the auxiliary atoms are excited, we observe that the point
of $\mathcal{R}=0$ shifts significantly to the right as the
number of atoms increases. However, in the
Fig. {\ref{W}}(b), when all auxiliary atoms are at ground states, the situation
is opposite to Fig. {\ref{W}} (a). 
It is worth emphasizing that these points of $\mathcal{R}=0$ are exactly the effective frequencies $\omega_L^\prime$ as mentioned previously. In this way, the phenomena of Fig. {\ref{RT}} can also be understood as follows.
The auxiliary atoms can enhance the increasing or decreasing effect on the rectification factor based on the different auxiliary atomic states.

In our model, the $L$ atom is coupled with $N$ auxiliary atoms, which have
$2^N$ possible states. According to   (\ref{R1}), the states of
the auxiliary atoms have a significant influence on the rectification effect.
Especially when all the auxiliary atoms are at excited states or all are at ground states, the corresponding rectification factors will serve as the two bounds of the
rectification effect, respectively. This means no matter what states the
auxiliary atoms are in, the corresponding rectification factor is always between the two bounds. 
To give an intuitive illustration,  let's consider the case of three auxiliary atoms as an example.
These three atoms have a total of eight states,
corresponding to eight independent subspaces, i.e. $\vert ee%
e\rangle$, $\vert eeg\rangle$, $%
\vert ege\rangle$, $\vert egg%
\rangle$, $\vert gee\rangle$, $\vert ge%
g\rangle$, $\vert gge\rangle$, $%
\vert ggg\rangle$. Each independent subspace will
have a different effect on the rectification effect of the system. Considering the effective
frequencies $\omega_L^\prime$ we introduced earlier, we can express the
effective frequencies in these eight independent subspaces unambiguously as $\omega_L^\prime\vert_{\vert eee\rangle}$=$%
\omega_L+6g_{L1}$, $\omega_L^\prime\vert_{\vert eeg%
\rangle}$=$\omega_L+2g_{L1}$, $\omega_L^\prime\vert_{\vert eg%
e\rangle}$=$\omega_L+2g_{L1}$, $\omega_L^\prime\vert_{\vert e%
gg\rangle}$=$\omega_L-2g_{L1}$, $\omega_L^\prime\vert_{%
\vert gee\rangle}$=$\omega_L+2g_{L1}$, $%
\omega_L^\prime\vert_{\vert geg\rangle}$=$%
\omega_L-2g_{L1}$, $\omega_L^\prime\vert_{\vert gge%
\rangle}$=$\omega_L-2g_{L1}$, $\omega_L^\prime\vert_{\vert g%
gg\rangle}$=$\omega_L-6g_{L1}$. Without loss of
generality, let $g_{La}$=$g_{L1}$. Based on the eight effective frequencies, the states can be divided into four types. The first type shows that all auxiliary atoms are in excited states, i.e., $\vert ee%
e\rangle$; The second type is formed by the combination of two
excited states and one ground state, i.e., $\vert eeg\rangle$%
, $\vert ege\rangle$, and $\vert ge%
e\rangle$; The third type consists of a
combination of two ground states and one excited state, in contrast to the second
type, i.e., $\vert egg\rangle$, $%
\vert geg\rangle$, and $\vert gg%
e\rangle$; The fourth type 
is that all the auxiliary atoms are in the ground states, i.e., $\vert g%
gg\rangle$. In Fig. {\ref{three}}, we plot the rectification factors for the eight cases with the temperature $T_L$.
It is shown that the rectification effects are completely coincident for the states belonging to the same type, and the first and the fourth types correspond to the two bounds.
Finally, we would like to emphasize that although the auxiliary atomic states may
enhance or inhibit the rectification effect, if $\sum_{a=1}^N g_{La}=0$, the auxiliary atoms
will no longer affect the rectification, which is mentioned at the end of Sec. III.
 \begin{figure}
\centering
\includegraphics[width=8.3cm]{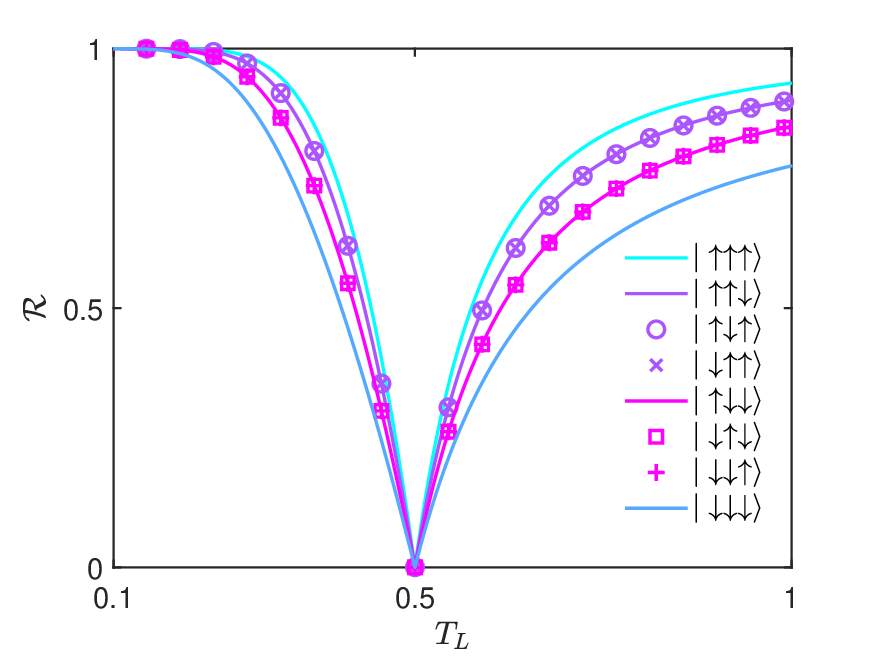} 
\caption{The rectification factor $\mathcal{R}$ varies with the temperature $%
T_L$. The different lines mark the different states of the auxiliary states. Here we take $\protect%
\omega_L=4$, $\protect\omega_R=2$, $\protect\omega_a$=2, $\protect\gamma%
=0.001$, $g_{LR}=0.1$, $g_{La}=0.1$, $T_R=0.5$.}
\label{three}
\end{figure}

Finally, we will also investigate the effect of superposition states on rectification. Let's take the example of a system
with one single auxiliary atom. When the $L$ atom is connected to one
auxiliary atom, the heat current between the $L$ atom and the reservoir is 
\begin{align}
\dot{Q}_L=p_1\dot{Q}_{L,1}+(1-p_1)\dot{Q}_{L,2},  \label{Q6}
\end{align}
where 
\begin{align}
\dot{Q}_{L,1}&=-4g_{LR}\Gamma_{1}^1,  \notag \\
\dot{Q}_{L,2}&=-4g_{LR}\Gamma_{2}^1,
\end{align}
represent the steady-state heat current of the system with the excited  and
ground-state auxiliary atom, respectively, and $p_1$ represents the fraction in
where the auxiliary atom is excited. The heat current is
proportional to the fraction $p_1$. 
Based on Eq. (\ref{Q5})  is equal to Eq. (\ref{Q6}), i.e., $\dot{Q}^\prime_L=\dot{Q}_L$, we get the critical fraction $p_c=\frac{{\dot{Q}_L^\prime}-\dot{Q}_{L,2}}{\dot{Q%
}_{L,1}-\dot{Q}_{L,2}}$.

Figure {\ref{P1}} shows that as $p_1$ changes, the heat current $\dot{Q}_L$
and the rectification factor $\mathcal{R}$ also change accordingly. The blue
solid line in the figure represents the heat current $\dot{Q}_L$, the red
solid line represents the rectification factor $\mathcal{R}$, and the red
dashed line represents the rectification factor $\mathcal{R}_0$ without
auxiliary atoms. Without the auxiliary atoms, the system has no independent
subspace. Therefore, the rectification factor $\mathcal{R}_0$ does not
change with the variation of $p_1$ and is a constant. When a $L$ atom is
connected to an auxiliary atom, the heat current $\dot{Q}_L$ increases
linearly with $p_1$, while $\mathcal{R}$ shows a
nonlinear change. When $p_1=p_c$, the auxiliary atom does not affect the
system. 
 Therefore, in the case of $\omega_L>\omega_R$, the auxiliary atom is beneficial to rectification when the fraction is greater than the critical fraction, i.e. $p_1>p_c$. When the system satisfies $\omega_L < \omega_R$ and the proportion of auxiliary atoms $p_1$ is lower than the critical value $p_c$, i.e. $p_1<p_c$, the auxiliary atoms can enhance the rectification.
\begin{figure}
\centering
\includegraphics[width=8.3cm]{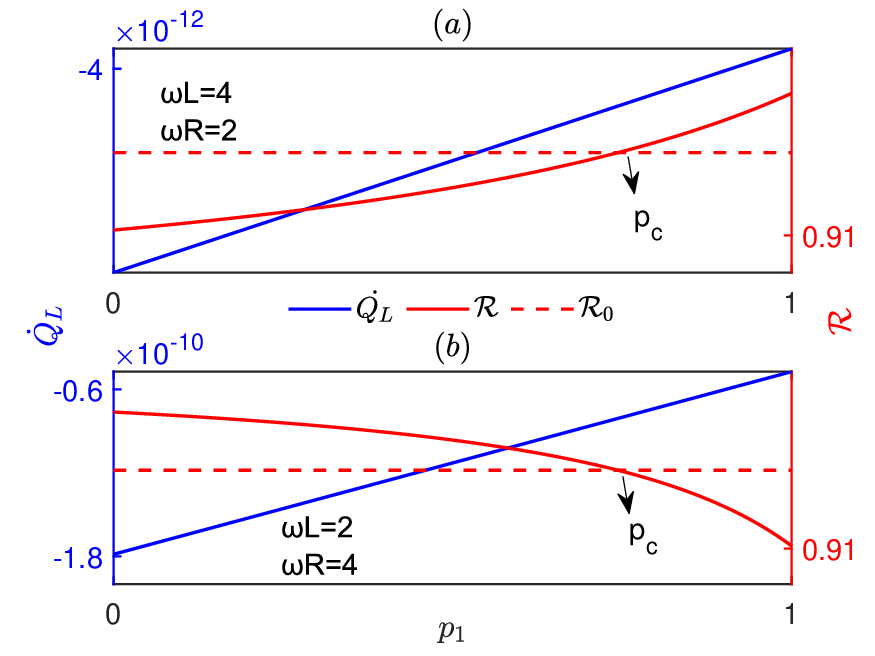} 
\caption{The heat current $\dot{Q}_L$ and the rectification effect $\mathcal{%
R}$ varies with $p_1$. The blue solid line represents heat current, the red
solid line represents rectification, and the red dashed line indicates
rectification without auxiliary atoms.  Here $\omega_1=2, \gamma=0.001, g_{LR}=0.1, g_{L1}=0.1, T_L=0.3, T_R=0.5.$}
\label{P1}
\end{figure}

In the above section,   there was no dissipation for the auxiliary atoms.  For integrity, we will only consider the effect of the dissipation.  Let's add an atom to the $L$ atom.  See Appendix {\ref{Appendix C}}
for the detailed derivation process. In Fig. {\ref{H}}, the blue line
indicates that the system is dissipative, while the red line represents the
non-dissipative state.  When the dissipation effect
is small, the rectification behaviour of the system is similar to that of
the system without dissipation, indicating that the
system can maintain its rectification characteristics under the condition of
weak dissipation, and the effect is close to that of the non-dissipative system situation.
This is also applicable to $N$ auxiliary atoms.

\begin{figure}
\centering
\includegraphics[width=8.3cm]{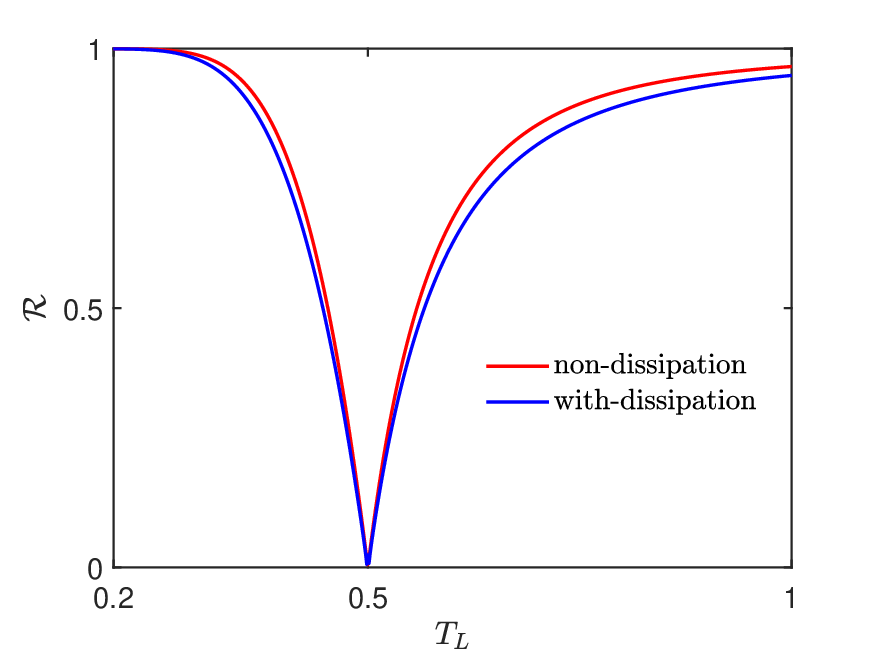} 
\caption{The rectification factor $\mathcal{R}$ varies with the temperature $T_L$ when the auxiliary atoms are dissipative. The blue (red) line indicates that the auxiliary atom is (or is not) connected to a
heat reservoir. $\protect\omega_L=4$, $\protect%
\omega_R=1$, $\protect\omega_1=5$, $\protect\gamma=0.001$, $\protect\gamma_1=%
\protect\gamma^2$, $g_{LR}=0.1$, $g_{L1}=0.1$, $T_R=0.5, T_1=0.8$.}
\label{H}
\end{figure}

Superconducting circuits can realize various thermodynamic devices, such as thermal diodes \cite{Sánchez_2017,Ordonez-Miranda2017,Senior2020,Díaz_2021}, thermal valves \cite{Liu2023}, and thermal transistors \cite{PhysRevB.101.184510,Yamamoto_2021}.  A quantum thermal diode star network can adopt a superconducting qubit architecture, with a central qubit at the core and peripheral qubits connected through capacitive or inductive coupling \cite{doi:10.1126/science.1107572}.  The central qubit dynamically regulates the coupling strength with the peripheral qubit by microwave modulation of its nonlinear inductance, forming a programmable star-coupled network \cite{doi:10.1126/science.1141324}. Drawing on the strong coupling technology between superconducting phase qubits and resonators, by using microwave pulses, the coupling strength and anisotropic parameters between the central spin and the surrounding spins can be dynamically regulated \cite{Hofheinz2008}, thereby achieving 
\begin{align}
H_I=J\sum_{i=1}^N(c_x S_0^{x}S_i^{x}+c_y S_0^{y}S_i^{y}+c_z S_0^z S_i^z),
\end{align} 
where  $J$ represents the antiferro-magnetic coupling constant, $S_0$ denotes the spin operator for the central spin,  $S_i$ corresponds to the spin operators of the surrounding outer spins, and the coefficients $c_x$, $c_y$, $c_z$ determine the nature of the spin coupling and are treated as controllable parameters \cite{Korkmaz2022}.

\section{Conclusion}
\label{section5}

We have designed a quantum thermal diode consisting of two coupled atoms,
each connected to an independent heat reservoir. One of the atoms is further
coupled to multiple auxiliary atoms. 
We find that there is a close relation between the states of the auxiliary atoms
and the heat current and the rectification performance of the system. For the case of $%
\omega_L>\omega_R$, the excited auxiliary atom can make
the best rectification effect. In contrast, when $\omega_L<\omega_R$, the
auxiliary atoms in the ground state have a positive impact on
rectification. The reason is that the effective frequency difference between the left
and the right system varies with the state of
the auxiliary atoms. The greater the frequency difference between the left
and right systems, the better the rectification effect. 
In addition, when the auxiliary atoms are in an arbitrary superposition state and $\omega_L>\omega_R$, we find that the rectification effect is enhanced when the parameter $p_1>p_c$. Conversely, when $\omega_L<\omega_R$, the system shows inverted behaviour, with enhancement occurring below the critical point, i.e., $p_1<p_c$. 
Finally, even if the auxiliary has weak dissipation, the impact on the system
rectification is small. The quantum thermal diode not only realizes high
efficiency rectification, but also provides a broad prospect for the
research and development of new quantum thermal management technology.

\section{Acknowledgements}

This work was supported by the National Natural Science Foundation of China under Grant No. 12175029.

\section{DATA AVAILABILITY}
	The data that support the findings of this article are not publicly available. The data are available from the authors upon reasonable request.

\begin{widetext}
\appendix 
\section{Evolution of off-diagonal elements} \label{Appendix A}
The equations for the dynamics of the off-diagonal elements are
\begin{align}
\nonumber
&\dot{\rho}_{i,j}=2J_{R}(+\omega_{R,1})\rho_{i+2^N,j+2^N}-[2J_{R}(-\omega_{R,1})+J_{L}(-\omega_{L,i})+J_{L}(-\omega_{L,j})]\rho_{i,j},\\
\nonumber
&\dot{\rho}_{i+2^N,j+2^N}=2J_{R}(-\omega_{R,1})\rho_{i,j}-[2J_{R}(+\omega_{R,1})+J_{L}(-\omega_{L,i+2^N})+J_{L}(-\omega_{L,j+2^N})]\rho_{i+2^N,j+2^N},\\
\nonumber
&\dot{\rho}_{i+2^{N+1},j+2^{N+1}}=2J_{R}(+\omega_{R,2})\rho_{i+2^{N+1}+2^N,j+2^{N+1}+2^N}-[2J_{R}(-\omega_{R,2})+J_{L}(\omega_{L,i})+J_L(\omega_{L,j})]\rho_{i+2^{N+1},j+2^{N+1}},\\
&\dot{\rho}_{i+2^{N+1}+2^N,j+2^{N+1}+2^N}=2J_R(-\omega_{R,2})\rho_{i+2^{N+1},j+2^{N+1}}-[2J_R(+\omega_{R,2})+J_L(+\omega_{L,i+2^N})+J_L(+\omega_{L,j+2^N})]\rho_{i+2^{N+1}+2^N,j+2^{N+1}+2^N}, 
\end{align}
where $i \in [1,2^N-1], j \in [i+1,2^N]$.

\section{Heat current and rectification factor}   \label{Appendix B}
\subsection{Heat current}
Using the vectorized population state $\vert \rho(t)\rangle$, according to Eq. (\ref{Q}), the steady-state heat current can be expressed as
\begin{align}
\nonumber
&\dot{Q}_{L,m}=\langle E_i\vert \mathcal{M}_L^{m}\vert\rho_{\mathit{ii}}^s\rangle,\\
&\dot{Q}_{R,m}=\langle E_i\vert \mathcal{M}_R\vert\rho_{\mathit{ii}}^s\rangle, \label{Q2}
\end{align}
where $\langle E_i\vert$  is the vector composed of the eigenvalues of the system energy.
According to Eq. (\ref{Q2}), the steady-state heat current of each independent subspace can be given
\begin{align}
\nonumber
&\dot{Q}_{L,m}=-(\omega_{m,m+2^{N+1}}\Gamma_{m,m+2^{N+1}}^L+\omega_{m+2^N,m+2^{N+1}+2^N}\Gamma_{m+2^N,m+2^{N+1}+2^N}^L),\\
&\dot{Q}_{R,m}=-(\omega_{m,m+2^{N}}\Gamma_{m,m+2^{N}}^R+\omega_{m+2^{N+1},m+2^{N+1}+2^N}\Gamma_{m+2^{N+1},m+2^{N+1}+2^N}^R). \label{Q3}
\end{align}
Substituting Eq.(\ref{state}) into $\Gamma_{i,j}$, we can obtain that the net transition rate is the same in each independent subspace, i.e., $\Gamma_{m,m+2^{N+1}}^L=-\Gamma_{m+2^N,m+2^{N+1}+2^N}^L=\Gamma_{m,m+2^{N}}^R=-\Gamma_{m+2^{N+1},m+2^{N+1}+2^{N}}^R\equiv\Gamma_m^N$. Furthermore, it was also discovered that $\omega_{m,m+2^{N+1}}-\omega_{m+2^N,m+2^{N+1}+2^N}=\omega_{m,m+2^{N}}-\omega_{m+2^{N+1},m+2^{N+1}+2^{N}}=4g_{LR}$. So Eq. (\ref{Q3}) is rewritten as
\begin{align}
\nonumber
&\dot{Q}_{L,m}=-4g_{LR}\Gamma_{m,m+2^{N+1}}^L,\\
&\dot{Q}_{R,m}=4g_{LR}\Gamma_{m,m+2^{N}}^R.  \label{Q4}
\end{align}
\subsection{Rectification factor}
Based on the heat current expressions of each independent subspace, the heat rectification effects of each subspace can be calculated. Assuming that the reverse heat current is greater than the positive heat current, Eq. (\ref{R}) can be rewritten as
\begin{align}
\mathcal{R}=1-\frac{\dot{Q}_L^f}{\dot{Q}_L^r}. \label{R3}
\end{align}
On the contrary, the numerator and the denominator are interchanged. According to Eq. (\ref{Q4}), the positive heat current and reverse heat current can be obtained as 
\begin{align}
\nonumber
\dot{Q}_L^f=\frac{2g_{LR}e^{-\frac{2g_{LR}(T_L+T_R)}{T_LT_R}}(e^{\frac{4g_{LR}}{T_L}}-e^{\frac{4g_{LR}}{T_R}})\gamma}{{\cosh}(\frac{2g_{LR}}{T_R})+{\cosh}(\frac{\omega_{R}}{T_R})+{\cosh}(\frac{2g_{LR}}{T_L})
+{\cosh}(\frac{\omega^\prime_{L}}{T_L})+{\sinh(\frac{2g_{LR}}{T_R}){\sinh}\frac{\omega^\prime_L}{T_L})}
+{\sinh}(\frac{2g_{LR}}{T_L}){\sinh}(\frac{\omega_{R}}{T_R})},\\
\dot{Q}_L^r=\frac{2g_{LR}e^{-\frac{2g_{LR}(T_R+T_L)}{T_RT_L}}(e^{\frac{4g_{LR}}{T_R}}-e^{\frac{4g_{LR}}{T_L}})\gamma}{{\cosh}(\frac{2g_{LR}}{T_L})+{\cosh}(\frac{\omega_{R}}{T_L})+{\cosh}(\frac{2g_{LR}}{T_R})
+{\cosh}(\frac{\omega^\prime_{L}}{T_R})+{\sinh(\frac{2g_{LR}}{T_L}){\sinh}\frac{\omega^\prime_L}{T_R})}
+{\sinh}(\frac{2g_{LR}}{T_R}){\sinh}(\frac{\omega_{R}}{T_L})}. \label{R4}
\end{align}
Substituting Eq. (\ref{R4}) into Eq. (\ref{R3}), we obtain the rectification factor.

\section{Rectification with non-dissipation}   \label{Appendix C}
The Hamiltonian of the system and the environment in the system is 
\begin{align}
\nonumber
H_{S1}=&\frac{1}{2} \omega_{L} \sigma^{z}_{L}+\frac{1}{2} \omega_{R} \sigma^{z}_{R}+\frac{1}{2}\omega_{1} \sigma^{z}_{1}+g_{LR}  \sigma^{z}_{L} \sigma^{z}_{R}+g_{L1}  \sigma^{z}_{L} \sigma^{z}_{1},\\
H_{E1}=&H_{E}+\sum_k\omega_{1 k}b_{1 k}^\dagger b_{1 k}.
\end{align}
Here, the subscript $\text{``1''}$ is used to identify the specific case where $N=1$.
The interactions between the environment and the system are
\begin{align}
H_{SE1}=H_{SE}+\sum_kf_{1 k}\sigma_1^x(b^\dagger_{1 k}+b_{1 k}).
\end{align}
The total Hamiltonian of the system is  $H=H_{S1}+H_{E1}+H_{SE1}$.

To study the master equation of the system, the eigenvalues of the system need to be solved. After a brief calculation, the eigenvalue of the system is 
\begin{align}
\nonumber
E_1&=\frac{1}{2}(+\omega_L+\omega_R+\omega_1)+g_{LR}+g_{L1}, &E_2=\frac{1}{2}(+\omega_L+\omega_R-\omega_1)+g_{LR}-g_{L1},\\
\nonumber
E_3&=\frac{1}{2}(+\omega_L-\omega_R+\omega_1)-g_{LR}+g_{L1}, &E_4=\frac{1}{2}(+\omega_L-\omega_R-\omega_1)-g_{LR}-g_{L1},\\
\nonumber
E_5&=\frac{1}{2}(-\omega_L+\omega_R+\omega_1)-g_{LR}-g_{L1}, &E_6=\frac{1}{2}(-\omega_L+\omega_R-\omega_1)-g_{LR}+g_{L1},\\
\nonumber
E_7&=\frac{1}{2}(-\omega_L-\omega_R+\omega_1)+g_{LR}-g_{L1}, &E_8=\frac{1}{2}(-\omega_L-\omega_R-\omega_1)+g_{LR}+g_{L1}.\\
\end{align}

Therefore, we can further calculate the eigenoperator of the system and finally obtain the eigenoperator as follows. 
The eigenoperators and the eigenfrequencies of the system are denoted by $V_{\mu l}$ ($l=1,\cdots,4$) and $\omega_{\mu l}$, respectively.  Concise calculations can give
\begin{align}
V_{L1} &= \vert5\rangle\langle 1\vert, & \omega_{L1} &= \omega_L+2g_{LR}+2g_{L1}, & V_{L2} &= \vert6\rangle\langle 2\vert, & \omega_{L2} &= \omega_L+2g_{LR}-2g_{L1}, \nonumber \\
V_{L3} &= \vert7\rangle\langle 3\vert, & \omega_{L3} &= \omega_L-2g_{LR}+2g_{L1}, & V_{L4} &= \vert8\rangle\langle 4\vert, & \omega_{L4} &= \omega_L-2g_{LR}-2g_{L1}, \nonumber \\
V_{R1} &= \vert3\rangle\langle 1\vert, & \omega_{R1} &= \omega_R+2g_{LR}+2g_{L1}, & V_{R2} &= \vert4\rangle\langle 2\vert, & \omega_{R2} &= \omega_R+2g_{LR}-2g_{L1}, \nonumber \\
V_{R3} &= \vert7\rangle\langle 5\vert, & \omega_{R3} &= \omega_R-2g_{LR}+2g_{L1}, & V_{R4} &= \vert8\rangle\langle 6\vert, & \omega_{R4} &= \omega_R-2g_{LR}-2g_{L1}, \nonumber \\
V_{11} &= \vert2\rangle\langle 1\vert, & \omega_{11} &= \omega_1+2g_{LR}+2g_{L1}, & V_{12} &= \vert4\rangle\langle 3\vert, & \omega_{12} &= \omega_1+2g_{LR}-2g_{L1}, \nonumber \\
V_{13} &= \vert6\rangle\langle 5\vert, & \omega_{13} &= \omega_1-2g_{LR}+2g_{L1}, & V_{14} &= \vert8\rangle\langle 7\vert, & \omega_{14} &= \omega_1-2g_{LR}-2g_{L1}, \nonumber
\end{align}
and the eigenoperator $V_{\mu l}$ corresponds to  the positive eigenfrequency.

After the auxiliary atoms are connected to the thermal reservoir, the coefficient matrix in $\vert\dot{\rho}\rangle=\mathcal{M}\vert\rho\rangle$ is $\mathcal{M}=\mathcal{M}_L+\mathcal{M}_R+\mathcal{M}_1$, where
\begin{align}
\nonumber
\mathcal{M}_L &= 2 \begin{pmatrix}
-J_L(-\omega_{L1})   &  0 & 0 & 0 &  J_L(+\omega_{L1})  &  0 & 0 & 0 \\
0   &  -J_L(-\omega_{L2}) & 0 & 0 & 0  &   J_L(+\omega_{L2})  & 0 & 0 \\
0   &  0 & -J_L(-\omega_{L3}) & 0 & 0  &   0  & J_L(+\omega_{L3}) & 0 \\
0   &  0 & 0 & -J_L(-\omega_{L4}) & 0  &   0  & 0 & J_L(+\omega_{L4}) \\
J_L(-\omega_{L1})   &  0 & 0 & 0 &  -J_L(+\omega_{L1})  &  0 & 0 & 0 \\
0   &  J_L(-\omega_{L2}) & 0 & 0 & 0  &   -J_L(+\omega_{L2})  & 0 & 0 \\
0   &  0 & J_L(-\omega_{L3}) & 0 & 0  &   0  & -J_L(+\omega_{L3}) & 0 \\
0   &  0 & 0 & J_L(-\omega_{L4}) & 0  &   0  & 0 & -J_L(+\omega_{L4}) 
\end{pmatrix},\\
\nonumber
\mathcal{M}_R &= 2 \begin{pmatrix}
-J_R(-\omega_{R1}) & 0 &J_R(+\omega_{R1}) & 0 & 0 & 0 & 0 & 0\\
0 & -J_R(-\omega_{R1}) &0 &J_R(+\omega_{R1}) & 0 & 0 & 0 & 0\\
J_R(-\omega_{R1}) & 0 &-J_R(+\omega_{R1}) &0 & 0 & 0 & 0 & 0\\
0 & J_R(-\omega_{R1}) &0 &-J_R(+\omega_{R1}) & 0 & 0 & 0 & 0\\
0 & 0 & 0 & 0 & -J_R(-\omega_{R2})  & 0  &J_R(+\omega_{R2}) &0\\ 
0 & 0 & 0 & 0 & 0  & -J_R(-\omega_{R2})  &0  &J_R(+\omega_{R2})\\
0 & 0 & 0 & 0 & J_R(-\omega_{R2})   & 0  &-J_R(+\omega_{R2}) &0\\
0 & 0 & 0 & 0 & 0   & J_R(-\omega_{R2})  &0 &-J_R(+\omega_{R2})
\end{pmatrix},\\
\mathcal{M}_1 &= 2 \begin{pmatrix}
-J_1(-\omega_{11})  & J_1(+\omega_{11}) & 0 & 0 & 0 & 0 & 0 & 0\\
J_1(-\omega_{11})   & -J_1(+\omega_{11}) & 0 & 0 & 0 & 0 & 0 & 0\\
0  & 0 & -J_1(-\omega_{11}) & J_1(+\omega_{11}) & 0 & 0 & 0 & 0\\
0  & 0 & J_1(-\omega_{11}) & -J_1(+\omega_{11}) & 0 & 0 & 0 & 0\\
0 & 0 & 0 & 0 & -J_1(-\omega_{12}) &J_1(+\omega_{12}) & 0 & 0 \\
0 & 0 & 0 & 0 & J_1(-\omega_{12}) &-J_1(+\omega_{12}) & 0 & 0 \\
0 & 0 & 0 & 0 & 0 & 0 & -J_1(-\omega_{12}) & J_1(+\omega_{12})\\
0 & 0 & 0 & 0 & 0 & 0 & J_1(-\omega_{12}) & -J_1(+\omega_{12})
\end{pmatrix}.
\end{align}
Given the coefficient matrix $\mathcal{M}$, $\rho$ can be solved.

\end{widetext}

\bibliography{new.bib}

\end{document}